\documentclass[11pt,a4paper]{article}
\usepackage[hyperref]{acl2020}
\usepackage{times}
\usepackage{graphicx}
\usepackage{graphics}
\usepackage{latexsym}
\usepackage{seqsplit}
\usepackage{indentfirst}

\usepackage{microtype}

\aclfinalcopy 

\title{Thistle: A Vector Database in Rust}

\author{Brad Windsor \thanks{All authors are equal contributors and NYU-affiliated.} \\
  \texttt{bw1879@nyu.edu} \\\And
  Kevin Choi \footnotemark[1]\\
  \texttt{kc2296@nyu.edu} \\}

\date{}

\begin{document}
\maketitle
\begin{abstract}
We present Thistle \footnote{All code is open-sourced at \seqsplit{https://github.com/bwindsor22/thistle}}, a fully functional vector database. Thistle is an entry into the domain of latent knowledge use in answering search queries, an ongoing research topic at both start-ups and search engine companies. We implement Thistle with several well-known algorithms, and benchmark results on the MS MARCO dataset. Results help clarify the latent knowledge domain as well as the growing Rust ML ecosystem.
\end{abstract}

\section{Introduction \& Related Work}
Knowledge representation is an unsolved problem in Natural Language Processing, a study of how machines ought to store information gathered from text such that it is useful for downstream tasks. 

Broadly, one category of knowledge representation is explicit; words are kept in a human understandable format and explicit connections are made. Examples include Prolog rules banks, Knowledge Bases like WikiData \cite{vrandevcic2014wikidata}, or SQL tables. 

Another is implicit knowledge, keeping formats easier for machines, especially neural networks, to make use of. Examples include Word2Vec \cite{goldberg2014word2vec}, ELMo \cite{peters2018deep}, and BERT \cite{devlin2018bert}. Such numerical representations are often made by pre-training networks on large text corpuses. 

Most open-source search engines rely on an explicit knowledge representation by matching term counts between an input query and each sample corpus match. Specifically, both ElasticSearch and Vespa are based on BM-25 \cite{robertson1995okapi}, a modification of the TF-IDF counts. This is quite easy for humans to understand; a search for ``Blue Armadillo" will be more likely to return documents matching ``Armadillo" than ``Blue" because ``Armadillo" is a rare word.

However, search engines are increasingly beginning to look for ways to incorporate implicit knowledge. Google recently began using BERT to revise search results \cite{nayak_2019}, NBoost\footnote{https://github.com/koursaros-ai/nboost} is a neural booster for ElasticSearch, and PineconeDB\footnote{http://pinecone.io} forgoes the BM-25 entirely and operates only on vectors.

Thistle seeks to operate only in the implicit domain, doing search only with sentence embeddings from neural networks. We look at several ways of doing this comparison, and measure database accuracy and runtime.

As a secondary goal, Thistle is implemented in Rust to help evaluate machine learning libraries there. Rust is a new programming language attracting recent support from Linux, Google, and Facebook as a low-complexity low-level language.

\section{Methods}
\subsection{Approach}
Thistle's approach can be outlined as follows: pieces of sentence-level text are turned into embeddings of size 768 via SBERT (Sentence-BERT), and then two exact nearest neighbor search algorithms (one using cosine distance and one using Euclidean distance) as well as three approximate nearest neighbor search algorithms (two using the proximity graph method and one using locality sensitive hashing) are provided to retrieve the top k results closest to a given query. The user selects which algorithm to use at runtime. Details with regards to these algorithms are delineated below.

\begin{figure}[h]
\centerline{\includegraphics[scale=0.28]{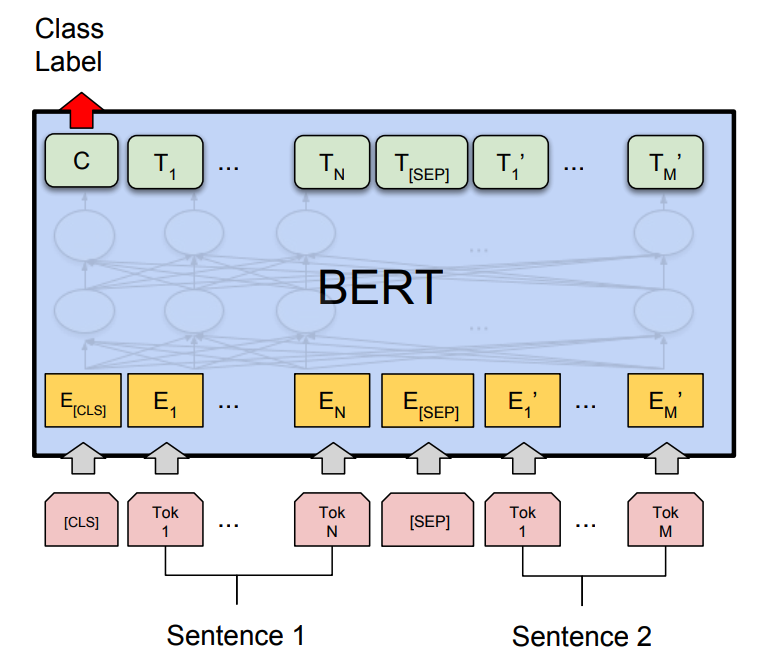}}
\caption*{BERT's architecture}
\end{figure}

\subsection{SBERT}
Applying the bidirectional training of the attention-based Transformer model to language modeling, BERT has achieved state-of-the-art results since 2018. Despite its adaptability to various tasks in NLP, one disadvantage of BERT is that no independent sentence embedding is computed such that one has to use, as a proxy, either the average of the BERT output layer (a la average word embeddings) or the output of the [CLS] token (included in the beginning of every sentence in BERT's training process). Unfortunately, both of these not only fall short in the task of similarity comparison and correspondingly information retrieval (which Thistle tackles), but also suffer from substantial computation costs from a training perspective.

\begin{figure}[h]
\centerline{\includegraphics[scale=0.6]{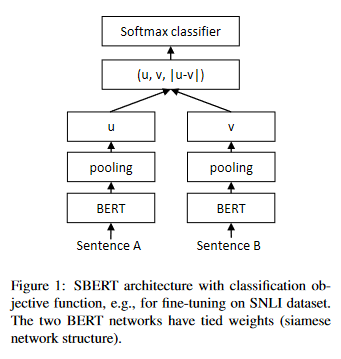}}
\end{figure}

To address the above limitation of BERT, SBERT is designed to output a stand-alone sentence embedding using BERT. The idea is to add a pooling operation to the output of BERT, followed by siamese and triplet networks for fine-tuning purposes. The pooling operation is in fact a hyperparameter. Three options include using the output of the aforementioned [CLS] token, computing the mean of all output vectors, and computing a max-over-time of the output vectors. All three of these options are provided by Thistle. SBERT substantially outperforms BERT, Universal Sentence Encoder, and others on key benchmarks, making it the basis for our work \cite{reimers2019sentencebert}.

\subsection{Search Algorithms}
\subsubsection{Iterative}
As a basic benchmark, we explore the cumbersome option of comparing an input embedding to every embedding in the corpus a la exact kNN. This is implemented in Rust without library support.

\subsubsection{HNSW}
\begin{figure}[h]
\centerline{\includegraphics[scale=0.4]{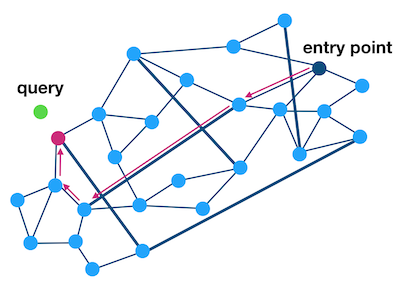}}
\caption*{An example of a proximity graph}
\end{figure}
The proximity graph method helps perform approximate kNN by conceptually turning a set of data points into a graph whose nodes represent the data points respectively and are connected if and only if particular geometric distance requirements between two nodes are met. The basic algorithm can proceed in a greedy manner as follows. Given a query point, we begin the search by starting at some enter point (either random or predefined) as our initial base node, compute the distance between the query node and all the neighbors of the enter point, choose the neighbor that is closest to the query, update the base node to be that neighbor, and greedily repeat the process by computing the distance between neighbors of the updated base node and the query until some stopping condition is met (e.g. upper bound on the number of iterations).

While NSW (navigable small world) is an example of the proximity graph method utilizing navigable graphs (i.e. graphs with polylogarithmic scaling of the number of hops during the greedy traversal with respect to the network size), HNSW, short for hierarchical navigable small world, extends it by introducing a multi-layer structure comprising a hierarchical set of more than one proximity graphs for nested subsets of data points.

\begin{figure}[h]
\centerline{\includegraphics[scale=0.45]{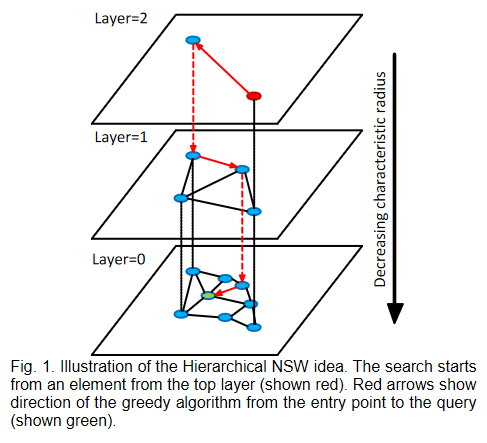}}
\end{figure}

As a simple example for intuition purposes, imagine we start with 8 data points, all of which are included in the ``base" (layer 0) proximity graph (which is precisely the only graph used in non-HNSW cases of the proximity graph method). Then we build another proximity graph (call it layer 1) using only a subset of the nodes from the layer 0 graph. Lastly, we can build a layer 2 proximity graph using only a subset of the nodes from layer 1. The number of layers can of course be much higher in practice and is a hyperparameter. The idea is that we can traverse the graph starting from the highest layer (as opposed to the base layer) and travel downwards given a query such that we are essentially able to ``skip" potentially unnecessary intermediary iterations of the greedy process, promoting efficiency. It is for this reason that Thistle leverages HNSW as opposed to other forms of the proximity graph method. While Thistle provides two distance metrics (Euclidean and cosine) for HNSW, it can be easily configured to support other metrics as well. Overall, HNSW accounts for 2 out of 3 approximate kNN methods that Thistle offers.

\subsubsection{LSH}
Locality sensitive hashing (LSH) represents yet another approximate kNN method supported by Thistle. It relies on the fact that data points that are similar in nature (i.e. close to each other) have a higher probability of getting hashed to the same bucket, which is untrue for general hashing (where hash collisions are minimized not maximized). Consequently, this technique can indeed be used for approximate kNN. Given a query, we simply hash it to some bucket using LSH and consider the data points in that bucket while keeping track of the algorithm's probability guarantees. It is possible to explore and configure such guarantees on Thistle via hyperparameters.

\begin{figure}[h]
\centerline{\includegraphics[scale=0.5]{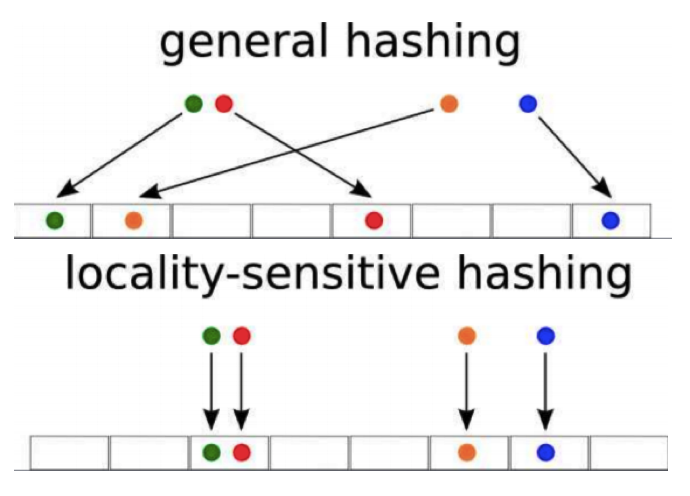}}
\caption*{Locality sensitive hashing (LSH)}
\end{figure}

\subsection{RUST}
Rust \cite{matsakis2014rust} is a new programming language that exceeds or comes within 5\% of C++ on many important benchmarks \cite{languages} like the n-body problem or digits of Pi. It is memory- and thread-safe, has no garbage collection, and is supported by a robust set of tools including excellent compiler error messages.

The website arewelearningyet.com describes machine learning in Rust as ``ripe for experimentation, but the ecosystem isn't very complete." We found this to be true; we iterated through three libraries to find an HNSW implementation that worked for us, and with LSH, we fixed an open issue \footnote{https://github.com/ritchie46/lsh-rs/issues/9} so the library would work again.

Our Rust implementation features a common interface as laid out by Traits, with load and query operations implemented for each of the database objects.

\section{Results}
\subsection{Evaluation}
The MS Marco dataset \cite{bajaj2018ms} is used for evaluation. This dataset is composed of anonymized questions sampled from Bing's search logs and separated into query and result tuples. 

Evaluation began by downloading and cleaning the dataset, removing special characters. For each of the 5 databases instances, the following evaluation was run: 
\begin{itemize}
    \item Insert the full result corpus into the database.
    \item For each (query, result) tuple in the corpus, run the query text on the database. If the result is the expected result, mark as correct.
\end{itemize}

Total time is taken as time to insert plus time to query, and the experiment is run for N = 100, 1,000, and 10,000 entries. See figures at end.

\subsection{Analysis}
\textbf{Trends} \newline
Several trends are immediately obvious: as dataset size increases, accuracy goes down, and runtime goes up.

As expected, a point by point comparison of the query and output through all data points produces the highest accuracy. Libraries such as LSH and HNSW are approximate algorithms meant to speed lookups. At the lowest dataset size, a straight-through iteration of all data is fastest.

Although LSH passed a local unit test, it had very poor accuracy across dataset sizes. This became clear in testing; as soon as more than a few words changed, LSH had difficulty making the match. Some tuning of hyperparameters might improve this result, as LSH takes as input the number of projections and hash tables. 

\textbf{Assessment} \newline
Overall, results bring significant sobriety to the idea of a fully vector-oriented search database.

 One concern is time; the HNSW Euclidean benchmark at 10,000 data points took 11 hours to run, hardly practical for a production application. Over 99\% of this time was in running SBERT, which is significantly slower than alternatives like Universal Sentence Encoder. Even as benchmark results for implicit representations improve, it is important to still do work to ensure representations can be computed in a reasonable time.
 
 Another concern is accuracy; at N=10,000, we see accuracies in the range of 30-40\%. One way in which this might be improved would be to use Neural Ranking on top of the BM-25 algorithm; only use neural networks to refine the top K results. A more immediate gain might be to constrain the results to passages that only fit within the SBERT token size, though this is also a constraint: additional code is needed to represent varying passage sizes with BERT-family models, though BM-25 handles this easily.

\section{Conclusion}

Overall Thistle is a successful instance of a purely latent knowledge-based database written in Rust. The project succeeded in benchmarking several Rust libraries and included a merged pull request on LSH. We were able to compare implementations and measure runtime and accuracy, though both would need improvement for a production system. Future work might include benchmarking results against existing explicit-knowledge databases (such as ElasticSearch), or exploring the use of latent knowledge to enhance searches in BM-25 databases. 
\newline
\newline
\newline
\newline
\newline
\newline
\begin{figure}[h]
\centerline{\includegraphics[scale=0.65]{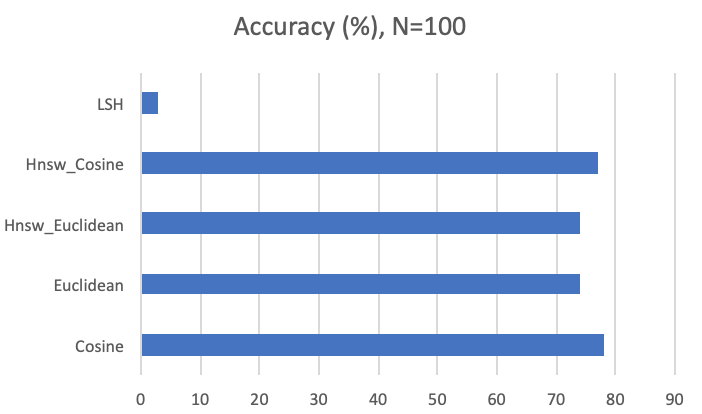}}
\end{figure}
\begin{figure}[h]
\centerline{\includegraphics[scale=0.65]{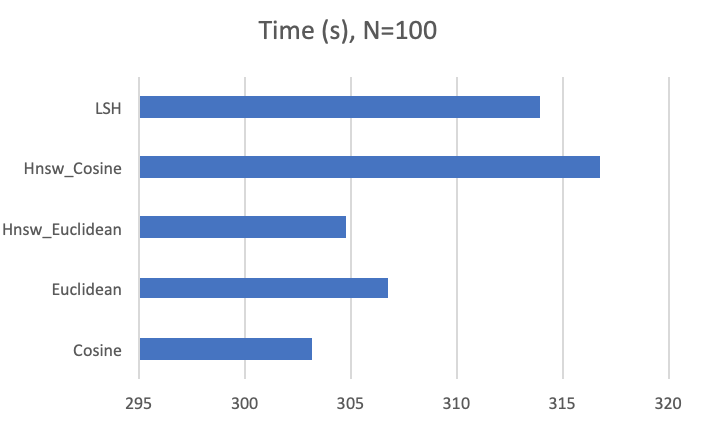}}
\end{figure}
\begin{figure}[h]
\centerline{\includegraphics[scale=0.65]{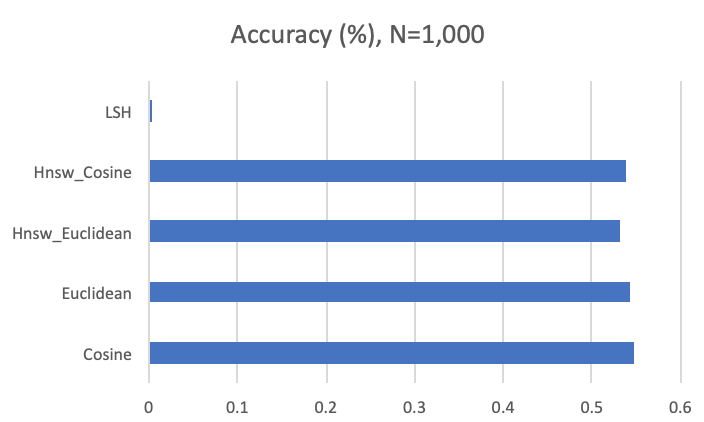}}
\end{figure}
\begin{figure}[h]
\centerline{\includegraphics[scale=0.65]{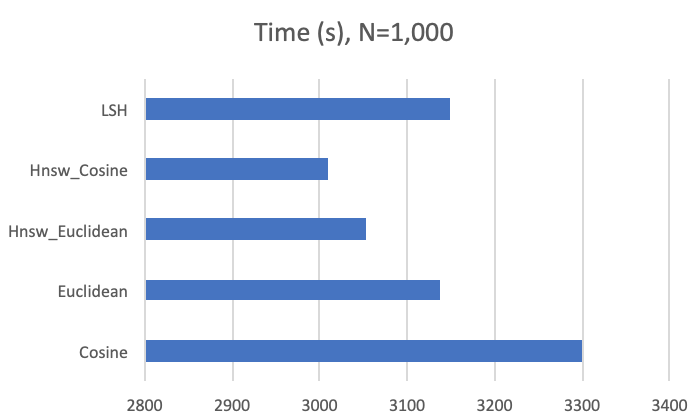}}
\end{figure}
\begin{figure}[h]
\centerline{\includegraphics[scale=0.65]{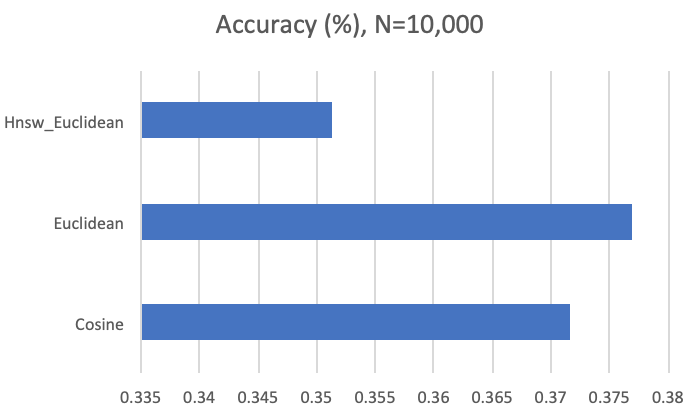}}
\end{figure}
\begin{figure}[h]
\centerline{\includegraphics[scale=0.65]{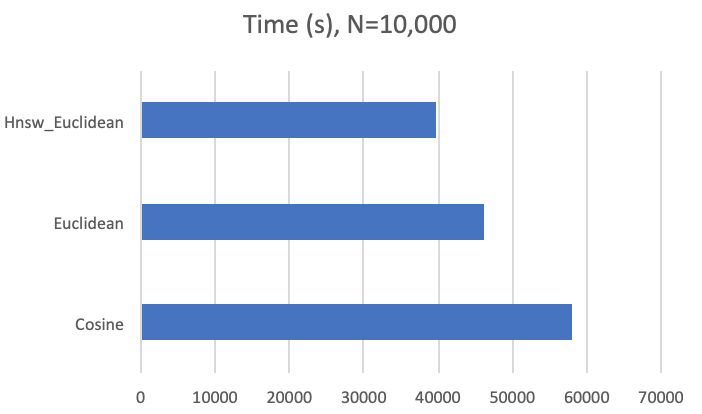}}
\end{figure}

\bibliography{anthology,acl2020}
\bibliographystyle{acl_natbib}

\end{document}